\documentclass[aps,pra,preprint,showpacs,showkeys,nofootinbib]{revtex4-1}
\usepackage{graphicx}
\usepackage{natbib}
\usepackage{dcolumn}
\usepackage{bm}
\usepackage[utf8]{inputenc}
\usepackage{color}

\newcommand{\ket}[1]{\left| #1 \right>}

\begin{document}

\title{Atom-mediated effective interactions between modes of a bimodal cavity}
\author{F.~O. Prado}\author{F.~S. Luiz}\altaffiliation[Currently at ]
{Departamento de F\'{\i}sica, Universidade Federal de S\~ao Carlos. S\~ao Carlos-SP, Brazil}
\author{J.~M. Villas-B\^{o}as}\author{A.M. Alcalde}\author{E.~I. Duzzioni}\author{L. Sanz}\email{lsanz@infis.ufu.br}
\affiliation{Instituto de F\'{\i}sica, Universidade Federal de Uberl\^{a}ndia, 38400-902,
Uberl\^{a}ndia-MG, Brazil}

\begin{abstract}
We show a procedure for engineering effective interactions between
two modes in a bimodal cavity. Our system consists of one or more
two-level atoms, excited by a classical field, interacting with both
modes. The two effective Hamiltonians have a similar form of a
beam-splitter and quadratic beam-splitter interactions,
respectively. We also demonstrate that the nonlinear Hamiltonian can
be used to prepare an entangled coherent state, also known as
multidimensional entangled coherent state, which has been pointed
out as an important entanglement resource. We show that the
nonlinear interaction parameter can be enhanced considering $N$
independent atoms trapped inside a high-finesse optical cavity.
\end{abstract}
\pacs{42.50.Ct,42.50.Pq,03.67.Bg,}
\keywords{Cavity Quantum Electrodynamics, nonlinear interaction, Entanglement}
\maketitle

\section{Introduction}
\label{sec:intro} Cavity Quantum Electrodynamics (CQED) is an ideal
scenario for research on fundamentals of quantum theory and quantum
information. Achievements on both, the construction of high quality
cavities and control of atom-field interactions, are associated with
the use of the entanglement properties for the successful generation
of quantum states of light, such as Einstein-Podolski-Rosen (EPR),
Schr\"{o}dinger cat and Fock states~\cite{RevCQED}. An important
consequence of the high experimental control in CQED was the
successful reconstruction of quantum states of light prepared inside
a high quality cavity~\cite{Bertet02,Deleglise08}. This procedure
allows, for example, the observation of decoherence process of a
Schr\"{o}dinger cat-like state, through the analysis of snapshots of
the Wigner function~\cite{Wigner32,Wigner84}. The same setup was
also used to reconstruct the Wigner function of Fock states with
more than one photon~\cite{Deleglise08}. Recently, CQED setups has
been used in order to study three-photon correlations~\cite{Koch11},
the apparition of electromagnetically induced transparency using
Rubidium~\cite{Mucke10} and Cesium~\cite{Kampschulte10} single atoms
and quantum jumps~\cite{Reick10,Khuda09}.

A particular CQED experimental setup could include a bimodal cavity.
In this kind of device, two bosonic modes with different
polarizations are prepared inside the cavity~\cite{Raus01}. In the
context of quantum information theory, a bimodal cavity is
interesting because the additional mode acts as a third photonic
qubit (besides the atom and the first cavity mode qubits), opening
new possibilities for implementation of quantum information
protocols~\cite{Messina03}. Potential applications of bimodal
cavities have been analyzed in some recent works. Those include the
implementation of quantum logic gates~\cite{Dong09} and generation
of entangled states~\cite{Gonta09,Gonta08}. Entanglement between the
two modes of a superconducting cavity was experimentally
demonstrated~\cite{Raus01}, where a maximally entangled state was
created.

Schr\"{o}dinger cat states can be generated by interaction between
atoms and the electromagnetic field confined in a high-quality
cavity. CQED schemes are used to prepare a superposition of two
packages, as the experiment reported by Del\'{e}glise \textit{et.
al}~\cite{Deleglise08}. A different approach to produce those states
is to use a nonlinear
Hamiltonian~\cite{Yurke86,Gerry102,Gerry99,Agarwal97,BanerjiPRA98}.
This method involves Kerr-like Hamiltonians and the superposition
states are created from the evolution of initial coherent states.
The ``size" of the superpositions is limited by the value of the
nonlinear parameter, which could be low~\cite{Glancy08}.

One of the features of CQED is the ability of manipulating physical
parameters in order to sculpt an effective interaction. From the
theoretical point of view, this ability can be explored by following
the procedure proposed by James and
co-workers~\cite{James00,James07,James10}. This well-established
method is used for the obtention of effective Hamiltonians, which
govern the dynamics of the system for a specific choice of physical
parameters on the exact Hamiltonian. Recent applications of this
method include the proposal of robust preparation of atomic W
states~\cite{Yang11}, the generation of NOON states in cavities
connected by an optical fiber~\cite{Can11} and the implementation of
entangling gates for two logical qubits in decoherence-free
subspaces~\cite{Feng09}.

In this work, based in our experience~\cite{Prado06,Werlang08}, we
use the method of Refs.~\cite{James00,James07,James10} to engineer
two effective Hamiltonians using the interaction of a two-level atom
with a bimodal cavity and laser fields. One is a CQED version of a
beam-splitter, the other is a quadratic beam-splitter Hamiltonian.
The generation of the proposed effective interactions opens
interesting possibilities such as interferometry using CQED, similar
to the atomic linear and nonlinear interferometry developed with
Bose-Einstein condensates~\cite{Oberthaler10}, and new schemes of
quantum state engineering and quantum information processing.
Concerning quantum state engineering, we demonstrate that one of the
potential applications of the effective quadratic beam-splitter is
to produce entangled generalized coherent states~\cite{Birula68}.
The entangled coherent state (ECS), also known as multidimensional
entangled coherent state, was first discussed by Tombesi and
Mecozzi~\cite{Tombesi86} and Sanders~\cite{Sanders92}. More
recently, van Enk proposed its generation using a Kerr medium and
analyzed the dynamics of entanglement~\cite{Enk03}. Other
theoretical proposals consider its creation by using
ions~\cite{Solano01} and CQED~\cite{Zou05} experimental setups.
Finally, we also show that the nonlinear interaction parameter can
be amplified by considering $N$ independent neutral atoms
interacting with the cavity modes.

We organized this paper as follows: In Section~\ref{sec:effective}
we obtain both effective Hamiltonians in the context of CQED by
considering a single atom interacting with classical and quantum
fields of light. The generation of ECS is presented in Section~\ref
{sec:dynamics}. In Section~\ref{sec:manyatoms}, we show how to
amplify the effective nonlinear coupling between the cavity modes
using a system composed of $N$ neutral atoms trapped in an optical
cavity. A discussion about experimental feasibility is contained at  section~\ref{sec:expfea}.
In section~\ref {sec:summary} we present our conclusions and
perspectives.

\section{Engineering the effective Hamiltonians}
\label{sec:effective} In this section, we show how to generate
effective two-modes Hamiltonians of CQED system. We consider two
cavity modes (mode A and B) with orthogonal
polarizations~\cite{Turchette95,Raus01,Duan04} interacting with an
atom prepared in an excited state. We consider a two-levels atom
with transition frequency $\omega _{0}$ between the ground
($\left\vert g\right\rangle $) and excited ($\left\vert
e\right\rangle $) states. The parameters $\lambda_{a,b}$ describe
the interaction between atom and cavity modes A and B with
frequencies $\omega_{a,b}$ respectively. The two-level atom also
interacts with a resonant classical field with Rabi frequency
$\Omega $. The full Hamiltonian can be written as ($\hbar =1$)
\begin{equation}
H=H_{0}+H_{\mathrm{I}},
\label{eq:iniH}
\end{equation}
where
\begin{eqnarray*}
H_{0} &=&\omega _{a}a^{\dagger }a+\omega _{b}b^{\dagger }b+\frac{\omega _{0}
}{2}\sigma _{z}, \\
H_{\mathrm{I}} &=&\left( \lambda _{a}a+\lambda _{b}b+\Omega e^{-i\omega
_{0}t}\right) \sigma _{eg}+\mathrm{H.c..}
\end{eqnarray*}
Here, $H_{0}$ describes a non-interacting system, where the
orthogonal polarization modes A and B of the cavity are associated
with the annihilation operators $a$ and $b$, respectively, and the
atomic operator is given by $\sigma _{z}=\left\vert e\right\rangle
\left\langle e\right\vert -\left\vert g\right\rangle \left\langle
g\right\vert$. The term $H_{\mathrm{I}}$ describes the atom-cavity,
and atom-classical field interactions. The atomic operator $\sigma
_{eg}=\left\vert e\right\rangle \left\langle g\right\vert$ describes
the promotion from ground to excited state.

At follows, we assume a cavity with degenerate modes
$(\omega=\omega_a=\omega_b)$ with equal coupling parameter to the
atomic transition ($\lambda=\lambda_a=\lambda_b$). Both condition
can be satisfied with well designed cavity, and, in this way, the
Hamiltonian can be written in the interaction picture as
\begin{equation}
H_{\mathrm{int}}=H_{\mathrm{cav}}+H_{\mathrm{cef}}  \label{eq:Hint}
\end{equation}
with
\begin{eqnarray}
H_{\mathrm{cav}} &=&\lambda\left(a+b\right)e^{i\Delta t} \sigma _{eg}+\mathrm{H.c.},  \nonumber \\
H_{\mathrm{cef}} &=&\Omega \sigma _{eg}+\mathrm{H.c.},\nonumber
\end{eqnarray}
where $\Delta=\omega _{0}-\omega$ is the detuning of the cavity
modes from atomic transition frequency. Assuming large detunings, so
that $\Omega\ll|\Delta|$ and
$|\Delta|\gg\sqrt{\overline{n}_{i}}|\lambda|$ ($i = a, b$), where
$\overline{n}_{i}$ is the mean number of photons in the $i$-th
cavity mode, $H_{\mathrm{cav}}$ presents fast oscillating time
dependence, which allows us to apply the \textit{effective
Hamiltonian approach} proposed in
references~\cite{James00,James07,James10}. From the high harmonic
disturbance of $H_{\mathrm{cav}}$, we can determine the dynamical
evolution by considering an averaged density matrix in a time
resolution which eliminates the high-frequency feature explicitly.
This averaging procedure preserves all relevant information about
the quantum system by inferring an effective Hamiltonian, and its
validity was discussed in details in Ref.~\cite{James10}.

Applying such procedure to the Hamiltonian (\ref{eq:Hint}) we obtain
\begin{equation}
 H_{\mathrm{int}}\simeq H_{\mathrm{cef}}+\chi(a^{\dag}a+b^{\dag}b+a^{\dag}b+ab^{\dag})\sigma_{z} + 2\chi\sigma_{ee},
\label{eq:Hef1}
\end{equation}
where $\chi\equiv\frac{\lambda^{2}}{\Delta}$. Notice that the second
term in Hamiltonian~(\ref{eq:Hef1}) can be interpreted as a
dispersive interaction between atom and cavity~\cite{Savage90}, as
the detuning $\Delta$ is large enough to avoid direct atomic
transitions. If the classical field is turned off ($\Omega=0$) and
the system is prepared as
\[
\left\vert \Psi (0)\right\rangle =\left\vert e\right\rangle \left\vert \psi
_{\mathrm{field}}(0)\right\rangle,
\]
the evolution of cavity states will be governed by the effective Hamiltonian written as
\begin{equation}
H_{\mathrm{BS}}=\chi(a^{\dag}a+b^{\dag}b+a^{\dag}b+ab^{\dag}),
\label{eq:hbs}
\end{equation}
with, in this case, $\chi\equiv\frac{\lambda^{2}}{\omega}$. This
effective Hamiltonian is similar to those obtained in
Ref.\cite{Prado06}: because the lack of a phase factor multiplying
the terms $a^{\dag}b$ and $ab^{\dag}$, it is interesting to notice
that the form of the above effective Hamiltonian has the same effect
of a $50/50$ beam splitter Hamiltonian over the cavity states. The
action of a beam splitter interaction is well known: it entangles
non classical field states, such as Fock and squeezed states, while
coherent and thermal states are not entangled~\cite{KimBS02}.

At follows, we will show how to engineer a nonlinear effective
interaction. Using the unitary transformation
$U=e^{-iH_{\mathrm{cef}}t}$, we can write the
Hamiltonian~(\ref{eq:Hef1}) in the rotating frame with Rabi
frequency $\Omega $ as
\begin{eqnarray}
H_{\mathrm{rf}} &=&U^{\dagger }H_{\mathrm{int}}U-H_{\mathrm{cef}}  \nonumber
\\
&=&\chi O\left( \sigma _{+-}e^{i2\Omega t}+\mathrm{H.c.}\right) ,
\label{eq:beam}
\end{eqnarray}
where
\begin{equation}
O\equiv a^{\dagger }a+b^{\dagger }b+a^{\dagger }b+ab^{\dagger }+
\mathbf{1}, \label{eq:operO}
\end{equation}
and $\sigma_{+-}=\left\vert +\right\rangle \left\langle -\right\vert $ is an atomic operator defined in the new basis
\begin{equation}
\left\vert \pm \right\rangle =\frac{\left\vert e\right\rangle \pm \left\vert
g\right\rangle }{\sqrt{2}}.  \label{eq:eapm}
\end{equation}
Imposing that $\Omega \gg \overline{n_{i}}\chi
,\overline{n_{i}}\sqrt{\overline{n_{j}}+1}\chi $ ($i,j=a,b$) and
applying again the same approach of
Refs.~\cite{James00,James07,James10} for Hamiltonian
(\ref{eq:beam}), we find the effective Hamiltonian
\begin{equation}
H_{\mathrm{NL}} =\frac{\chi ^{2}}{2\Omega }O^{2}\left( \sigma _{++}-\sigma _{--}\right) .
\label{eq:hnl}
\end{equation}
which is the nonlinear bosonic effective interaction between cavity
modes desired except by which is given by the term $\left( \sigma
_{++}-\sigma _{--}\right)/2\Omega$. This last term can be easily
eliminated from the dynamics by carefully choosing the atomic
initial state. Notice for instance that $\ket{+}$ and $\ket{-}$ are
eigenstates of Hamiltonian (\ref{eq:hnl}). Those states can be
experimentally created by applying a $\pi /2$ pulse of a classical
microwave field in an atom initially in the ground state $\ket{g}$
\cite{RevCQED}. Choosing the initial state of the atom-cavity after
this atom state preparation as
\begin{equation}
\left\vert \Psi (0)\right\rangle =\left\vert +\right\rangle \left\vert \psi
_{\mathrm{field}}(0)\right\rangle,  \label{eq:acis}
\end{equation}
the evolution ruled by Eq.~(\ref{eq:hnl}) is given by
\begin{equation}
\left\vert \Psi (t)\right\rangle =\left\vert +\right\rangle e^{-i\mu
O^{2}t}\left\vert \psi _{\mathrm{field}}(0)\right\rangle ,  \label{eq:aces}
\end{equation}
where $\mu =\frac{\chi ^{2}}{2\Omega }=\frac{ \lambda ^{4}}{2\Delta ^{2}\Omega }$ is the nonlinear coupling. This result shows that it is possible to build an effective interaction between both cavity fields as long as the conditions for dispersive interaction between atom and cavity are fulfilled and the atom is prepared in one of the states $\ket{+}$ or $\ket{-}$. In this case the effective quadratic beam-splitter (QBS) Hamiltonian, with one atom, is then given by
\begin{equation}
H_{\mathrm{QBS}}\simeq \mu O^{2}.  \label{eq:Hfoa}
\end{equation}
Here, as the operator $O^2$ depends on the square of the beam splitter interaction, it will entangle a product of coherent states. The generation of both effective interactions, Eq.~(\ref{eq:hbs}) and Eq.~(\ref{eq:Hfoa}), open interesting possibilities about interferometry using CQED, similar to the atomic linear and nonlinear interferometry developed with Bose-Einstein condensates~\cite{Oberthaler10}.

\section{Generation of ECS}
\label{sec:dynamics} In this section, we are particularly interested
in the creation of entangled superpositions of more than two
coherent states or ECS. In the context of CQED, Zou \textit{et
al.}~\cite{Zou05} proposed the creation of this kind of entagled
state also considering a bimodal cavity, following a probabilistic
procedure which implies that the field state is obtained after the
measurement of the atomic state. Also, it requires the passage of
several atoms, in order to increase the number of products of
coherent states. At follows, we demonstrate how to produce ECS
following a deterministic procedure, exploiting the dispersive
effective interaction between atom and cavity. We also show that it
is necessary only a passage of one atom, which can be useful in
order to control the effects of dephasing and decoherence processes.

To produce the ECS, we consider that both cavity modes are Glauber coherent states
\[
\ket{\psi_{\mathrm{field}}(0)} =\left\vert \alpha ,\beta
\right\rangle .
\]
These states are produced by the injection of two small coherent
fields oscillating in perpendicular directions with classical
amplitudes $\alpha $ (mode A) and $\beta $ (mode
B)~\cite{Glauber63}. Then, we explore the dynamics of the bimodal
cavity, ruled by the QBS Hamiltonian (\ref{eq:Hfoa}). The evolved
state of the field inside the cavity is given by
\begin{equation}
\left\vert \psi _{\mathrm{field}}(t)\right\rangle =e^{-iH_{\mathrm{QBS}}t}\left\vert \psi _{\mathrm{field}}(0)\right\rangle.
\label{eq:dynafield}
\end{equation}
As shown in the Appendix~\ref{sec:app}, the evolved state at times $ t_{g}=\frac{\pi }{2\mu }\frac{r}{s}=\tau_{\mu}\frac{r}{2s}$ is written as
\begin{equation}
\left\vert \psi \left( t_{g}\right) \right\rangle
=\sum_{p=0}^{j-1}a_{p}^{(r,s)}\left\vert \alpha _{f}(p)\right\rangle \otimes
\left\vert \beta _{f}(p)\right\rangle ,  \label{eq:cavcats}
\end{equation}
where $r$ and $s$ are prime numbers, $\left\vert \alpha _{f}(p)\right\rangle
$ and $\left\vert \beta _{f}(p)\right\rangle $ are coherent states, and
\begin{eqnarray}
\alpha _{f}(p) &=&2e^{-i\left( \mu t_{g}+\pi \frac{p}{j}\right) }\left[
\alpha \sin {\left( \theta _{p}\right) }-\beta \cos {\left( \theta
_{p}\right) }\right] ,  \nonumber \\
\beta _{f}(p) &=&2e^{-i\left( \mu t_{g}+\pi \frac{p}{j}\right) }\left[
\alpha \cos {\left( \theta _{p}\right) }-\beta \sin {\left( \theta
_{p}\right) }\right] ,  \nonumber \\
a_{p}^{(r,s)} &=&\frac{1}{j}\sum_{q=0}^{j-1}e^{-i\pi \frac{r}{s}q^{2}+2\pi i
\frac{p}{j}q},  \nonumber \\
\theta _{p} &=&\mu t_{g}+\pi \frac{p}{j}.  \label{parameters}
\end{eqnarray}
The expression above can be described as an entangled superposition
of coherent states. The number of terms on the sum depends on $j$,
which is fixed by the condition
\begin{equation}
j=\left\{
\begin{array}{ll}
2s & \mbox{if $r$ and $s$ are odd,} \\
s & \mbox{if $r$ is even and $s$ odd or vice versa.}
\end{array}
\right.  \label{eq:numbercs}
\end{equation}
The nonlinear terms in the effective Hamiltonian (\ref{eq:Hfoa}) are
the mechanics behind the formation of the superpositions. We also
observe exchange of photon population, which are connected with the
oscillatory functions of expression of $\alpha _{f}(p)$ and $\beta
_{f}(p)$. A particular case of Eq.~(\ref{eq:cavcats}) is obtained by
considering the initial state
\[
\left\vert \psi _{\mathrm{field}}(0)\right\rangle =\left\vert \alpha
,0\right\rangle ,
\]
which represents a specific experimental condition when a coherent
state is produce in the mode A, while the mode B remains empty. We
can check that the evolved state at times $t_{g}$ is still a
superposition with the same form of Eq.~(\ref{eq:cavcats}) but with
different amplitudes $\alpha _{f}(p)$ and $\beta _{f}(p)$, as can be
verified with Eqs. (\ref{parameters}).

Wigner functions are quasi-probability functions associated with
symmetric ordering of operators, which are equivalent to the density
matrix and are used to represent both, quantum superpositions and
statistical mixtures~\cite {Wigner32,Wigner84}. The Wigner function
can be obtained experimentally by performing measurements which
permits the reconstruction of the density matrix coefficients
associated with a specific physical situation. In the context of
QCED, methods for the measurement of the Wigner function of the
electromagnetic field in a cavity was first proposed
theoretically~\cite {Davidovich97} and then used in order to check
the actual state of electromagnetic field inside the
cavity~\cite{Nogues00}. Recently, the complete reconstruction of
Fock and Schr\"{o}dinger cat-like states was reported, so it becomes
possible to obtain snapshots of the decoherence process~\cite
{Deleglise08}.

To illustrate the form of the ECSs produced by the QBS Hamiltonian,
we compute the Wigner function associated with one of the cavity
modes. To obtain the Wigner function of the mode A, we write the
general density matrix of evolved state at time $t_{g}$ as
\[
\rho =\left\vert \psi _{\mathrm{field}}(t_{g})\right\rangle \left\langle
\psi _{\mathrm{field}}(t_{g})\right\vert .
\]
Then, by tracing over the variables associated with mode B, we
obtain the reduced density operator $\rho _{a}$
\begin{eqnarray}
\rho _{a} &=&\sum_{p,p^{\prime }=0}^{j-1}a_{p}^{(r,s)}a_{p^{\prime }}^{\ast
(r,s)}e^{-\frac{1}{2}\left[ \left\vert \beta _{f}(p)\right\vert
^{2}+\left\vert \beta _{f}(p^{\prime })\right\vert ^{2}-2\beta _{f}(p)\beta
_{f}^{\ast }(p^{\prime })\right] }  \nonumber \\
&&\times \left\vert \alpha _{f}(p)\right\rangle \left\langle \alpha
_{f}(p^{\prime })\right\vert .  \label{eq:rhoa}
\end{eqnarray}
At this point, we use the definition of the Wigner function~\cite{Wigner84}
\begin{equation}
W\left( \gamma \right) =\frac{1}{\pi }\int {d^{2}\xi e^{\xi \gamma ^{\ast
}-\xi ^{\ast }\gamma }\mathrm{Tr}\left( e^{-\xi a^{\dagger }+\xi ^{\ast
}a}\rho _{a}\right) },
\end{equation}
where $\gamma \equiv q_{a}+ip_{a}$, being $\left( q_{a},p_{a}\right)
$ the canonical variables of position and momentum related the mode
A.

\begin{figure}[tbp]
\includegraphics[scale=0.4]{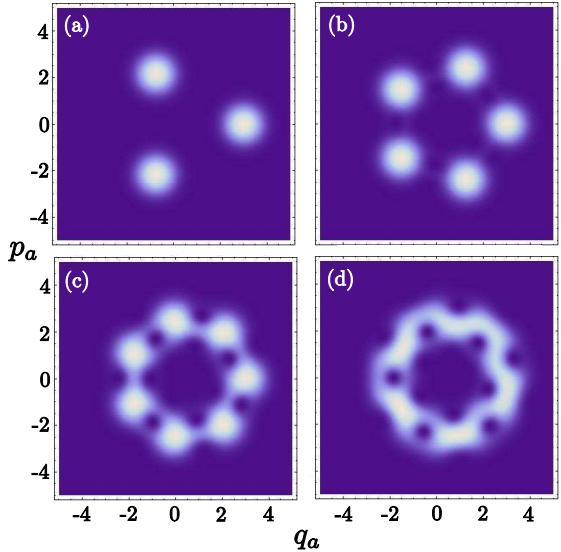}
\caption{(Color online) Wigner function for ECSs obtained for the
initial state with $\protect\alpha =3$ and $\protect\beta=2$
considering evolution times $t_g=\tau_{\mu}\frac{r}{2s}$ with $r=2$. (a) $t_g=\tau_{\mu}/3$ ($s=3$), (b) $t_g=\tau_{\mu}/5$ ($s=5$),
(c) $t_g=\tau_{\mu}/7$ ($s=7$) and (d) $t_g=\tau_{\mu}/11$ ($s=11$).} \label{fig:cats1}
\end{figure}

\begin{figure}[tbp]
\includegraphics[scale=0.4]{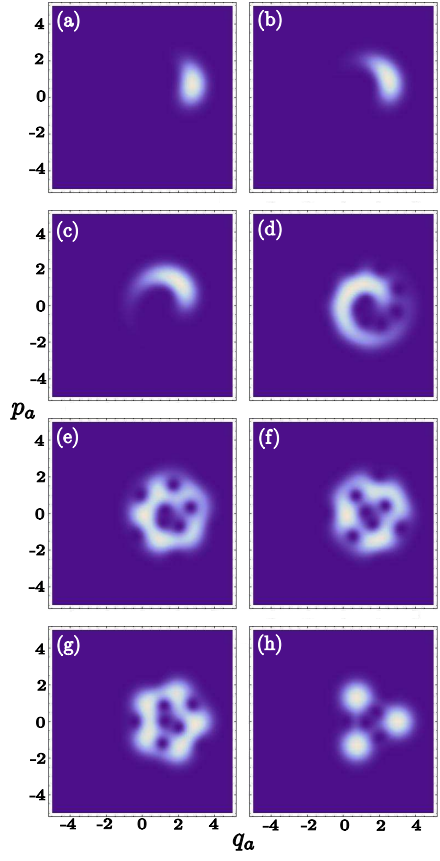}
\caption{(Color online) Snapshots showing the evolution of the
Wigner function associated with the cavity mode A for $\left\vert
\protect\psi _{ \mathrm{field}}(0)\right\rangle =\left\vert
3,0\right\rangle $ and $r=2$. Defining $t_{\protect\mu }\equiv
\protect\pi /\protect\mu $, we have: (a) $ t=t_{\protect\mu }/107$,
(b) $t=t_{\protect\mu }/61$, (c) $t=t_{\protect\mu }/37$, (d)
$t=t_{\protect\mu }/17$, (e) $t=t_{\protect\mu }/11$, (f) $t=t_{
\protect\mu }/7$, (g) $t=t_{\protect\mu }/5,$ and (h)
$t=t_{\protect\mu }/3$. } \label{fig:cats2}
\end{figure}

Figures~\ref{fig:cats1} and \ref{fig:cats2} shows the density plots
of the Wigner function for $\rho _{a}$. We are able to control the
number of packages, defined by the condition (\ref{eq:numbercs}),
which is shown in figure~\ref{fig:cats1}. The separation between the
packages depends on the initial mean value of photons inside the
cavity, given by $\left\vert \alpha \right\vert ^{2}+\left\vert
\beta \right\vert ^{2}$.
We can also use our analytical solution, Eq.~(\ref{eq:rhoa}), in
order to follow the dynamics at short times. In
figure~\ref{fig:cats2}, we plot snapshots of the Wigner function
considering $\alpha =3$, $\beta =0$, $r=2$ and decreasing values of
$s$, i.e., increasing values of evolution time $t_{g}$, which are
expressed as fractions of time scale $\tau_{\mu}=\pi/\mu $ parameter.
We can see that an initial coherent state at point $ \left(
q_{a},p_{a}\right) =\left( 3,0\right) $ starts to spread in phase
(Fig.\ref{fig:cats2}(a) to (c)) until the ``head'' meets the tail of
Wigner function. After that time, the state starts to interfere with
itself and it is possible to resolve different packages of
superposition.

\section{Amplifying the nonlinear coupling}
\label{sec:manyatoms} In this section, we demonstrate how to amplify
the nonlinear coupling on Hamiltonian (\ref{eq:Hfoa}) by using an
ensemble of $N$ identical neutral atoms. We consider all atoms with the same transition frequency $\omega_{0}
+ \epsilon$ between ground $\left\vert g\right\rangle $ and excited
$\left\vert e\right\rangle $ states. Each atom couples with both,
the classical field with Rabi frequency $\Omega $ and the
polarization modes in the cavity, with frequencies $\omega _{a}$ and
$\omega _{b}$. A sufficiently large interatomic separation is
considered so that the dipole-dipole interactions can be neglected.
In this case, we can describe the internal state of the atomic
assembly by the collective pseudo-spin operators written as
\[
J_{+}=\sum_{i=1}^{N}|e_{i}\rangle \langle
g_{i}|,\;J_{-}=\sum_{i=1}^{N}|g_{i}\rangle \langle e_{i}|,
\]
\begin{equation}
J_{z}=\sum_{i=1}^{N}(|e_{i}\rangle \langle e_{i}|-|g_{i}\rangle
\langle g_{i}|),
\end{equation}
which satisfy the angular momentum algebra. The Hamiltonian for $N$ atoms reads ($\lambda_a=\lambda_b=\lambda$)
\begin{eqnarray}
H^{N} &=&\omega _{a}a^{\dagger }a+\omega _{b}b^{\dagger }b+\frac{\omega _{0}
}{2}J_{z}+\epsilon J_{+}J_{-}\nonumber\\
&&+\lambda \left[ \left( a+b\right) J_{+}+\mathrm{H.c.}\right]+\Omega \left[ e^{-i\omega _{0}t}J_{+}+e^{i\omega _{0}t}J_{-}\right],
\label{eq:hamma}
\end{eqnarray}
where we consider the N atoms within a region of space whose linear
dimensions are smaller than the wavelength of cavity modes. Here,
the first four terms represent the free energy of the system, while
the fifth describes the interaction between the collection of atoms
with the cavity modes with coupling parameter given by $\lambda$. We
also consider the effect of a classical driving field on the
two-level atoms, described by the sixth term in
Eq.~(\ref{eq:hamma}). It is worth to note that the usual zero-point
energy reference of the two level atoms was changed with the
introduction of the $\epsilon$ parameter.

Following the same sequence of steps for obtaining the effective
Hamiltonian (\ref{eq:Hfoa}), we first go to the interaction picture.
The Hamiltonian (\ref{eq:hamma}) becomes
\begin{equation}
H_{\mathrm{int}}^{N}=H_{\mathrm{cav}}^{N}+H_{\mathrm{cef}}^{N}
\label{eq:HintN}
\end{equation}
with
\begin{eqnarray}
H_{\mathrm{cav}}^{N} &=&\lambda[(a+b)e^{-i\Delta t}J_{+} + \textbf{H}.c.] + \epsilon J_{+}J_{-}\nonumber \\
H_{\mathrm{cef}}^{N} &=&\Omega J_{+}+\Omega J_{-},\nonumber
\end{eqnarray}
so that $\Delta=\Delta _{a}\equiv \omega _{a}-\omega _{0}=\Delta
_{b}\equiv \omega _{b}-\omega _{0}$, i.e., the two-modes are
degenerate. The first term of $H_{ \mathrm{cav}}^{N}$ is the well
known Dicke Hamiltonian in the interaction picture. Again, we
consider that the frequencies of the cavity modes are far from
resonance with the atomic transition frequency so that the
dispersive condition $\left|\Delta\right| \gg
\sqrt{\overline{n}_{i}}\lambda $ is satisfied. Then, using the same
procedure of Ref.~\cite{James00,James07,James10} we obtain the
effective Hamiltonian
\begin{eqnarray}
H_{\mathrm{eff}}^{N} &\simeq &-2\chi \left( a^{\dagger }a+b^{\dagger }b+ab^{\dagger }+a^{\dagger
}b\right) J_{z}+H_{\mathrm{cef}}^{N},
\end{eqnarray}
where $\chi$ is the dispersive coupling defined previously and we
are using the condition $\epsilon = 2\chi$, just to remove the
effective shifts in all atomic excited states. The validity of the
effective Hamiltonian requires that $\Omega (\sim
\sqrt{\overline{n}_{i}}\lambda )\ll\left|\Delta\right|/N$. This
condition enables to disregard the influence of the classical
driving field on the bimodal dispersive interaction in according
with the numerical simulations from the Hamiltonian
(\ref{eq:hamma}).

Now we go to the rotating-frame, by using the unitary transformation
$ U\left( t\right) =e^{-i(J_{+}+J_{-})\Omega t}$,
obtaining
\begin{eqnarray}
H_{\mathrm{rf}}^{N} &\simeq &U^{\dag }H_{\mathrm{eff}}^{N}U-\dot{U}^{\dag }U
\\
&=&-\chi \left( a^{\dagger }a+b^{\dagger }b+ab^{\dagger }+a^{\dagger
}b\right) \left( \widetilde{J}_{+}(t)+\widetilde{J}_{-}(t)\right),\nonumber
\label{eq:Hosc}
\end{eqnarray}
where we have defined new collective atomic operators
\begin{eqnarray}
\widetilde{J}_{+}(t) &=&\sum_{i=1}^{N}|+_{i}\rangle \langle -_{i}|\exp
(i2\Omega t),  \nonumber \\
\widetilde{J}_{-}(t) &=&\sum_{i=1}^{N}|-_{i}\rangle \langle +_{i}|\exp
(-i2\Omega t),  \nonumber \\
\widetilde{J}_{z} &=&\sum_{i=1}^{N}(|+_{i}\rangle \langle
+_{i}|-|-_{i}\rangle \langle -_{i}|),
\end{eqnarray}
with $|\pm _{i}\rangle =\frac{1}{\sqrt{2}}(|e_{i}\rangle \pm
|g_{i}\rangle)$. Using again the effective Hamiltonian approach, we
obtain the effective interaction of many atoms with the cavity and
the classical field
\begin{eqnarray}
H_{\mathrm{ma}} &\simeq &\mu O^{2}\widetilde{J}_{z},
\label{eq:Hfmaan}
\end{eqnarray}
where
\begin{equation}
O \equiv a^{\dag}a + b^{\dag}b + a b^{\dag}+ a^{\dag}b .
\end{equation}
Consider that all atoms are prepared in the superposition state $
\ket{+_i}$ so the collective atomic state is
$\prod_{i=1}^{N}\ket{+_{i}}$. By using the eigenvalue relation
\[
\tilde{J}_{z} \prod_{i=1}^{N}\vert +_{i}\rangle = N \prod_{i=1}^{N}\vert +_{i}\rangle ,
\]
the evolved state associated with Hamiltonian (\ref{eq:Hfmaan}) is given by
\begin{eqnarray}
\vert\Psi(t)\rangle&=&\exp (-i \mu O^{2} \tilde{J}_{z}
t)\prod_{i=1}^{N}\ket{+_{i}}\ket{\psi_{\mathrm{field}}(0)}\nonumber\\
&=&\exp (-i
N \mu O^{2} t)\ket{\psi_{\mathrm{field}}(0)}
\prod_{i=1}^{N}\ket{+_{i}}, \label{eq:evolN}
\end{eqnarray}
which means that the dynamics of the modes inside the cavity depends on the amplified quadratic beam splitter (AQBS) Hamiltonian written as
\begin{equation}
H_{\mathrm{AQBS}}\simeq N\mu O^{2}.
\label{eq:Hfma}
\end{equation}
We can conclude that the coupling strength of the bimodal
Hamiltonian can be amplified by the factor $N$, when compared with
the one-atom case, Eq.~(\ref{eq:Hfoa}).

\begin{figure}[tbp]
\includegraphics[scale=0.5]{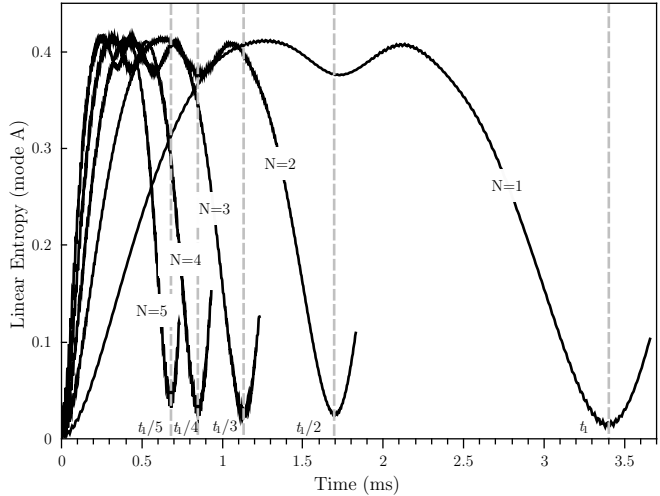}
\caption{Linear entropy for mode A as a function of time (in
microseconds) considering the evolution of initial state given by
$\prod_{i=1}^{N}\ket{+_{i}}\ket{\alpha,\beta}$ with
$\alpha=(q,p)=(1,0)$ and $\beta=(0,1)$ and different values of N.
The Hamiltonian parameters are $\Delta=12.5\lambda$,
$\Omega=\lambda$ with $\lambda=3\times 10^5$ Hz~\cite{RevCQED}.} \label{fig:delta}
\end{figure}

In order to check the validity of this amplification, we perform a
numerical calculation of linear entropy considering the exact
Hamiltonian (\ref{eq:HintN}) considering $N=1$ to $5$. The linear
entropy is a useful quantity which gives information about the
purity of the system. We are interested in the linear entropy for
the cavity mode described by operator $\hat{a}$ (mode A) defined as
\begin{equation}
\xi(t) = 1 - Tr_{a}\lbrace(\rho_{a}(t))^{2}\rbrace ,
\end{equation}
where $\rho_{a}(t)$ is the reduced density matrix of the cavity mode
A at time $t$. If $\xi=0$, the subsystem is pure and the state of
the system can be written as a direct product. To perform the
simulation, we consider the initial state as
$\prod_{i=1}^{N}\ket{+_{i}}\ket{\psi_{\mathrm{field}}(0)}=\prod_{i=1}^{N}\ket{+_{i}}\ket{\alpha,\beta}$
with $\alpha=(q,p)=(1,0)$ and $\beta=(0,1)$. The Hamiltonian
parameters are $\Delta=12.5\lambda$, $\Omega=\lambda$ and we use the value
$\lambda=3\times 10^5$ Hz from Ref.~\cite{RevCQED}. Figure~\ref{fig:delta} shows our results
for linear entropy of mode A as function of time considering $N=1$
to $5$ atoms. At the initial time the linear entropy is zero, in
agreement with the fact that the initial state is a direct product
($\ket{\alpha}\otimes\ket{\beta}\otimes\prod_{i=1}^{N}\ket{+_{i}}$).
The dynamics of linear entropy shows that the state of the
atom-cavity system could not be written as a direct product except
at the purification time, $t_1\sim\pi/\mu$. As we increase
the number of atoms, the purification time decrease following the
rule $t_N=t_1/N$. This is directly related with the effective
coupling which goes from $\mu$ (for one atom) to $N\mu$ (for $N$
atoms).

\section{Experimental feasibility}
\label{sec:expfea} In this section, we discuss some aspects about
current experimental feasibility of our proposal considering
different experimental setups of CQED~\cite{RevCQED,Koch11}. In the
experimental setup of Haroche \textit{et
al.}~\cite{Deleglise08,RevCQED,Raus01,Sayrin11}, Rydberg atoms
(rubidium) are coupled to a microwave high quality superconductivity
cavity. By considering the typical values of atom-cavity interaction
being $\lambda=2\pi \times 47$ kHz for experiments with $^{87}$Rb
and setting the detuning as $\Delta=2\pi\times 235$ kHz, we estimate
the value of effective frequency as $\chi= 2\pi\times 9.4$ kHz. In
that context, it is possible to perform a $\pi$-pulse operation
using Hamiltonian (\ref{eq:hbs}) at the time scale given by
$\tau_{\chi}\simeq\pi/\chi\simeq 0.05$ ms. Concerning the nonlinear
Hamiltonian (\ref{eq:Hfoa}) the coupling parameter is given by
$\mu=2\pi\times0.94$ kHz ($\Omega\sim\lambda$), which means that the
time required for a $\pi$-pulse is $\tau_{\mu}\simeq 0.5$ ms.
Entangled coherent states are created at lower times: in order to
create the ECS shown in Fig.~\ref{fig:cats1}, the time scale is
given by $t_g=\tau_{\mu}/11\simeq 0.045$ ms to $\tau_{\mu}/3\simeq
0.17$ ms. These times are smaller than the typical Rydberg atom
decay time ($\sim 30$ ms) and significatively smaller than the
decoherence time associated with cavity modes ($\sim 0.13$
s)~\cite{Deleglise08,RevCQED,Raus01}. In these experiments, the time
of interaction between atom and cavity depends on the velocity of
the atom ($100-600$ ms$^{-1}$) and varies between $100$ ns to $0.3$
ms~\cite{RevCQED}. The required times for the achievement of
beam-splitter Hamiltonian and the creation of ECS are both in this
time range but the realization of a complete  $\pi$-pulse due to the
nonlinear Hamiltonian is not.

The second experimental setup, used by Rempe \textit{et
al.}~\cite{Koch11}, consists of trapped two-level $^{85}$Rb atoms
(with atomic decay time $\sim 0.66$ $\mu$s) introduced in a small
ultra-high finesse optical cavity.  The atom-mode coupling is
stronger than the one mentioned above being $\lambda=2\pi\times 16$
MHz. The detuning between atomic transition and the cavity can be
controlled by an auxiliary laser. For $\Delta=2\pi\times 80$ MHz, we
estimate the effective beam-splitter coupling as $\chi=2\pi\times
3.2$ MHz with $\tau_{\chi}\simeq 0.16$ $\mu$s and the value of
nonlinear parameter is $\mu= 2\pi\times 0.32$ MHz which gives
$\tau_{\mu}\simeq 1.6$ $\mu$s ($\Omega\sim\lambda$). Thus, the
necessary evolution times, $t_g$, in order to create ECS as shown in
Fig.~\ref{fig:cats1}(b) and (d) are $0.32$ $\mu$s and $0.14$ $\mu$s,
respectively. The decoherence time scale of the optical cavity used
in this setup is given by $0.33$ $\mu$s, which favors both, the
implementation of the $\pi$-pulse with beam-splitter interaction and
the creation of ECS states but limits the implementation of
$\pi$-pulses with the nonlinear Hamiltonian.

In conclusion, the comparison between both experimental setups
points out that microwave cavity is a promising candidate to the
implementation of the one-atom scheme. Modifications on atomic
source or an auxiliar technique for slowing the atoms can be used in
order to explore all the advantages of the nonlinear effective
Hamiltonian. Another possibility is to use a continuous beam of
atoms, as those used in Ref.~\cite{Sayrin11}, so the nonlinear
interaction could be stabilized for the time required by the
operation. Nevertheless, although simultaneous interaction between
cavity and two atoms were reported~\cite{Osnaghi01}, the $N$-atoms
amplification could be difficult in this particular experimental
setup. Optical cavities, in contrast, are a promising system for the
implementation of our propose of amplification because neutral atoms
can be quasi-permanently trapped and the number of trapped atoms can
be increased one-by-one~\cite{Mucke10}. Another advantage is that
the atom-cavity interaction is a parameter that could be easily
controlled. The main problem in this setup is the decoherence of the
cavity field which we expect will be solved in the near future.

\section{Conclusions and perspectives}
\label{sec:summary} In this work, we use the effective Hamiltonian
approach~\cite{James00,James07,James10} in order to obtain two
effective interactions between the modes of a bimodal cavity,
Hamiltonians~(\ref{eq:hbs}) and~(\ref{eq:Hfoa}). By starting the
system state in a product of Glauber coherent states and for
specific times $t_{g}=\frac{\pi }{2\mu } \frac{r}{s}$, the nonlinear
Hamiltonian drags the system to a ECS. We are able to control the
number of packages manipulating either the time of evolution or
effective interaction parameter between quantum and classical fields
with the atomic system. Amplification of the nonlinear effective
coupling between the two-modes field, described by
Hamiltonian~(\ref{eq:Hfma}), can be obtained by considering a system
composed of $N$ two-level atoms trapped inside a bimodal
high-finesse optical cavity. We also discuss the experimental
feasibility of our proposal by checking the current value of
atom-cavity interaction considering both, microwave and optical
cavities. We estimate the values of effective coupling strengths,
$\chi$ and $\mu$, and the time scales associated with both, the
application of $\pi$-pulses, $t_{\chi}$ and $t_{\mu}$, and the
generation of entangled coherent states ($t_{\mathrm{ECS}}$). The
$\pi$-pulse with beam-splitter Hamiltonian and the generation of ECS
are possible in both scenarios. The implementation of a $\pi$-pulse
with nonlinear Hamiltonian (\ref{eq:Hfma}) requires a slightly
slower atom in the microwave scheme and a longer time of decoherence
in the optical setup.

Future works in this application includes the study of entanglement
properties associated with the nonlinear Hamiltonian and the effects
of decoherence on the entangled coherent states.

\section{Acknowledgments}
\label{sec:ack}
This work was supported by the Brazilian National
Institute of Science and Technology for Quantum Information
(INCT-IQ) and for Semiconductor Nanodevices (INCT-DISSE), CAPES,
FAPEMIG and CNPq.

\section{Appendix: Dynamics on cavity modes}
\label{sec:app}
Here, we briefly explain how to obtain the evolved
state associated with the QBS Hamiltonian (\ref{eq:Hfoa}). We can
rewrite Eq.~(\ref{eq:dynafield}) using the unitary transformation
$V=e^{\frac{\pi }{4 }\left( a^{\dagger }b-ab^{\dagger }\right) }$
defining the propagator $U(t)$ as follows
\begin{equation}
U(t)=e^{-iH_{\mathrm{oa}}t}=Ve^{-i\mu \left( 2b^{\dagger }b+1\right)
^{2}t}V^{\dagger },  \label{eq:propagator}
\end{equation}
so the evolved state takes the form
\begin{equation}
\left\vert \psi _{\mathrm{field}}(t)\right\rangle =U(t)\left\vert \psi _{
\mathrm{field}}(0)\right\rangle .
\end{equation}
We are interested in the dynamics when the initial state is a direct product of coherent states
\[
\ket{\psi _{\mathrm{field}}(0)} =\ket{\alpha
,\beta}=D\left( \alpha \right) D\left( \beta \right)\ket{0,0},
\]
where $D\left( \gamma \right)$ ($\gamma=(\alpha,\beta)$) is the
displacement operator: when working with unitary transformation $V$
and the product $D\left( \alpha \right) D\left( \beta \right)$, we
can use the identities:
\begin{eqnarray}
V^{\dagger }D\left( \alpha \right) D\left( \beta \right) V &=&D\left( \frac{
\alpha -\beta }{\sqrt{2}}\right) D\left( \frac{\alpha +\beta }{\sqrt{2}}
\right) ,  \nonumber \\
VD\left( \alpha \right) D\left( \beta \right) V^{\dagger } &=&D\left( \frac{
\beta +\alpha }{\sqrt{2}}\right) D\left( \frac{\beta -\alpha }{\sqrt{2}}
\right).  \label{eq:vdisv}
\end{eqnarray}
These expressions were used in order to obtain
Eq.~(\ref{eq:cavcats}). After the application of operator
$V^{\dagger }$ over initial state, we obtain
\begin{eqnarray}
\left\vert \psi _{\mathrm{field}}(t)\right\rangle &=&\hat{V}e^{-\frac{i}{
\hbar }\mu\left( 2b^{\dagger }b+1\right) ^{2}t}V^{\dagger
}\left\vert \alpha ,\beta \right\rangle  \nonumber \\
&=&\hat{V}\left\vert \frac{\beta +\alpha }{\sqrt{2}}\right\rangle \otimes
e^{-\frac{i}{\hbar }\mu\left( 2b^{\dagger }b+1\right)
^{2}t}\left\vert \beta _{v}\right\rangle .  \nonumber
\end{eqnarray}
with $\beta _{v}=\frac{\beta -\alpha }{\sqrt{2}}$. Expanding the
coherent state $\left\vert \beta _{v}\right\rangle $ in the Fock
basis on operator $ \hat{n}_{b}=b^{\dagger }b$, it is
straightforward to act with the second term of the
propagator~(\ref{eq:propagator}) on $\left\vert \beta
_{v}\right\rangle $ obtaining
\begin{eqnarray}
e^{-\frac{i}{\hbar }\mu\left( 2b^{\dagger }b+1\right)
^{2}t}\left\vert \beta _{v}\right\rangle &=&e^{-\frac{\left\vert \beta
_{v}\right\vert ^{2}}{2}}e^{-i\frac{\mu t}{\hbar }}\sum_{m}{
\frac{\left( \beta _{v}e^{-\frac{-4i\mu t}{\hbar }}\right) ^{m}}{
\sqrt{m!}}}  \nonumber \\
&&\times e^{-\frac{4i\mu m^{2}t}{\hbar }}\left\vert
m\right\rangle .  \label{eq:gcs}
\end{eqnarray}
This kind of superposition of Fock state is known as generalized
coherent state (GCS), which was introduced by Titulaer and
Glauber~\cite{Titulaer66}. At times given by $t_{g}=\frac{\pi
}{2\mu}\frac{r}{s}$, it is possible to rewrite the GCS state given
by Eq.~(\ref{eq:gcs}) as a superposition of coherent
states~\cite{Banerji101}
\[
e^{-\frac{i}{\hbar }\mu\left( 2b^{\dagger }b+1\right)
^{2}t}\left\vert \beta _{v}\right\rangle
=\sum_{p=0}^{l-1}a_{p}^{(r,s)}\left\vert \beta _{p}\right\rangle
\]
with $\beta _{p}=\beta _{v}e^{-2i\theta _{p}}$ and $\theta
_{p}=\frac{2\mu }{ \hbar }+\pi \frac{p}{l}$. Using the last result,
we write the evolved state as:
\[
\left\vert \psi _{\mathrm{field}}(t)\right\rangle
=\sum_{p=0}^{l-1}a_{p}^{(r,s)}\hat{V}\left\vert \frac{\beta +\alpha }{\sqrt{2
}}\right\rangle \left\vert e^{-2i\theta _{p}}\frac{\beta -\alpha }{\sqrt{2}}
\right\rangle .
\]
Finally, using the relations (\ref{eq:vdisv}), we arrive to Eq.~(\ref{eq:cavcats}).
\bibliographystyle{apsrev4-1}
\end{document}